\documentclass[aps,prb,twocolumn,superscriptaddress,floatfix]{revtex4}
\usepackage{graphicx}
\usepackage{epsfig}
\usepackage{float}

\newcommand{\be}{\begin{equation}}
\newcommand{\ee}{\end{equation}}
\begin{document}
\title{Electronic Structure of ZnCNi$_3$ }
\author{M. D. Johannes}
\affiliation{Department of Physics, University of California,
          Davis CA 95616}
\affiliation{Code 6391, Naval Research Laboratory, Washington, D.C.}
\author{W. E. Pickett}
\affiliation{Department of Physics, University of California,
          Davis CA 95616}

\begin{abstract} According to a recent report by Park {\it et al},
ZnCNi$_3$ is isostructural and isovalent to the superconducting (T$_c$
$\sim$ 8 K) anti-perovskite, MgCNi$_3$, but shows no indication of a
superconducting transition down to 2K.  A comparison of calculated electronic
structures shows that the main features of MgCNi$_3$, particularly the
van Hove singularity near the Fermi energy, are preserved in ZnCNi$_3$.  
Thus the reported lack of superconductivity in ZnCNi$_3$ is not explainable
in terms of T$_c$ being driven to a very low value by a small Fermi level 
density of states.  We propose that the lack of
superconductivity, the small value of the linear specific heat
coefficient $\gamma$ and the discrepancy
between theoretical and experimental lattice constants can all be
explained if the material is assumed to be a C-deficient $\alpha$-ZrCNi$_3$
similar to the analogous non-superconducting phase of MgCNi$_3$.

\end{abstract}

\maketitle

\section{Introduction}

The appearance of superconductivity \cite{THQH+01} near 8 K in the
Ni-rich perovskite, MgCNi$_3$, has stimulated much interest not only
because it is unusual in a compound that is primarily Ni, but because the
exact nature of the superconducting state and its microscopic origins are
still being debated.  Like the other unusual new superconductor, MgB$_2$,
it has so far resisted efforts to increase the critical temperature
significantly by chemical substitution.  Both Cu and Co doping on the Ni
site reduce the critical temperature (T$_c$), predictably due to band
effects (electron doping) in the former case and possibly due to spin
fluctuations in the latter \cite{Hay01, TGK+02}.  The transition
temperature can be raised by 1 K through Ni-site doping \cite{TGK+02}
with Fe, but this temperature occurs in MgCNi$_{3-x}$Fe$_x$ with x =
0.05, and any further doping again reduces T$_c$. Mg deficiencies or
excesses have some effect on the sharpness and onset of the
superconducting transition, but the optimal composition still results
\cite{Ren1,LiZhu} in a maximum T$_c$ of 8 K.  The superconductivity of
MgCNi$_3$ seems most sensitive to the carbon site occupancy.  Boron
doping on the carbon site \cite{Ren1} reduces T$_c$ for relative B/C
concentrations of up to 0.07 and eliminates superconductivity for any
greater concentration.  MgC$_x$Ni$_3$ with $x<$1.0 remains a cubic
perovskite but undergoes an isostructural transition \cite{Ren1, THQH+01}
to a smaller volume $\alpha$-phase that no longer superconducts.

Recently the synthesis of ZnCNi$_3$ has been reported by Park {\it et
al} \cite{M-P+}. Since ZnCNi$_3$ is very similar to MgCNi$_3$
structurally, and (as we will show) electronically, the lack of a
superconducting transition down to 2 K is quite unexpected.  
Understanding why superconductivity is seen in one compound but not the
other could be important in resolving remaining questions about the
unusual behavior of MgCNi$_3$.  The experimental data suggests
\cite{M-P+} that a strongly depressed density of states (DOS) (compared
to MgCNi$_3$) at the Fermi level (E$_F$) could be responsible for
pushing the transition temperature of ZnCNi$_3$ below 2 K.  The results
of a careful comparison of the electronic structure of the two
compounds are presented here, and the required lowering of the DOS is
shown to be absent.  Because the reported lattice constants differ by
4\%, we calculate the theoretical equilibrium lattice constants and
explore the effects of pressure on the electronic structure of each
material.  We find that discrepancies when compared to MgCNi$_3$ in
DOS, in lattice constant, and in observations of superconductivity can
be understood if the reported ZnCNi$_3$ samples are C-deficient 
$\alpha$-phase as are
MgCNi$_3$ samples with C deficiency. 

\section{Calculational Methods and Electronic Structure}

ZnCNi$_3$ has the typical ABO$_3$ cubic perovskite structure, but with
the oxygen atoms on the faces replaced by Ni atoms.  As Zn and Mg both
have a formal valency of 2$^+$, ZnCNi$_3$ is isovalent as well as
isostructural with MgCNi$_3$, both residing in space group 221
(Pm$\overline{3}$m). Calculations were carried out using Wien2k
\cite{wien}, a full-potential, augmented plane wave + local-orbital
method, and with the local density approximation (LDA) of Perdew and
Wang \cite{LDA5} to the exchange-correlation potential.  The density is
well-converged with 816 k-pts in the irreducible Brillouin zone. The
sphere radii used were 2.1 a.u. for Zn/Mg, and 1.72 a.u. for both C and
Ni; the Rkmax was set to 7.00. The experimental lattice constants were
used in the initial calculations for both MgCNi$_3$ (a = 3.81 $\AA$)
and ZnCNi$_3$ (a = 3.66 $\AA$).  Compression and expansion percentages
are given in terms of these experimental values.

The electronic structure of MgCNi$_3$ has been presented previously by
several groups \cite{DJS01, JHS01, SBDTJ01, RosPRL02}.  The dominant
feature is a remarkable, sharp van Hove singularity 65 meV below E$_F$,
which was traced to an extremely flat band around the M point
(=(1,1,0)$\frac{\pi}{a}$) of the Brillouin zone. The electronic
structure of ZnCNi$_3$ is very similar to that of MgCNi$_3$.  The sharp
peak just below the Fermi energy is still dominant, though it is
shifted slightly downward in energy by approximately 30 meV and has
broadened somewhat (Fig. 1, top panel).  In both compounds, the Ni ions
are two-fold coordinated with their nearest-neighbors, the co-planar C
ions.  Hybridization between Mg/Zn and Ni ions is very small,
consistent with the very similar electronic structures of the two
compounds.  The dispersion created by the two-dimensional bonding of
Ni-d and C-p orbitals is responsible \cite{RosPRL02, DJS01} for the
nearly dispersionless band centered on M.  In ZnCNi$_3$, the situation
is much the same, but the 4p states of the Zn ions, with which the Ni
ions are four-fold coordinated, do participate weakly in the bonding
states near the Fermi energy.  This weak bonding is three-dimensional,
accounting for the slightly increased dispersion around the M-point as
well as the lowered energy of the DOS peak.

The downward shift of the peak has the effect of reducing the DOS at
the Fermi energy (N(0)), with respect to that of MgCNi$_3$, by about 1
eV$^{-1}$, i.e. by about 20 \%.  This decrease relative to the Mg
compound is much less than what is necessary to account for the lack of
superconductivity through conventional BCS theory (see Section C).

\begin{figure}[t]
\hspace{-0.3 in} \includegraphics[width=3.8 in, angle=270]{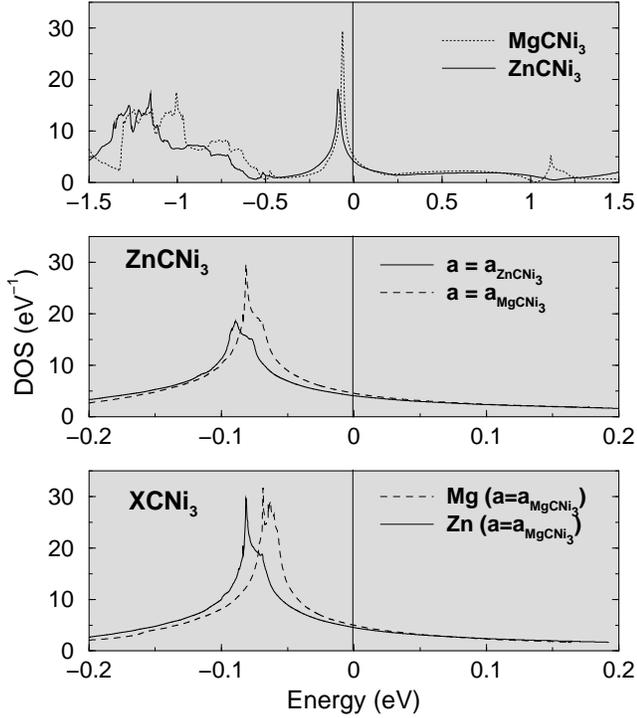}

\caption{ {\it Top panel:} ZnCNi$_3$ and MgCNi$_3$ at their reported
equilibrium lattice constants. {\it Middle panel:} ZnCNi$_3$ at its own
lattice constant and at that of MgCNi$_3$.  The effect of pressure is
rather small. {\it Bottom panel:} MgCNi$_3$ and ZnCNi$_3$ both at the
same
lattice constant (that of MgCNi$_3$).  Differences in electronic
structure
are larger than can be accounted for by pressure alone, though still
small.}
\end{figure}

\subsection{Pressure Dependencies and Bulk Modulus}
Expansion of the ZnCNi$_3$ lattice narrows the peak and brings it
nearer the Fermi level (Fig. 1 middle panel) that is, expansion makes
it more MgCNi$_3$-like.  This has the effect of raising the DOS at the
Fermi energy, but the change is very small even for fairly large
expansions.  In ZnCNi$_3$, an expansion of $\sim$ 12\% by volume caused
a change in the DOS of only 10\%.  MgCNi$_3$ seems to be even slightly
{\it less} sensitive than this.  A calculation of ZnCNi$_3$ at the
equilibrium lattice constant of MgCNi$_3$ shows, however, that the
differences in the electronic structures of the two compounds are due
to more than simply volume.  (See bottom panel of Fig. 1).  The
probable source of the small differences in electronic structure is
residual hybridization of Ni-d and Zn-p orbitals.  As the lattice is
expanded the overlap between these orbitals decreases, but does not
disappear completely. By looking at the orbitally-resolved character of
the flat band, we have observed that the amount of Zn-p character in
ZnCNi$_3$, although minor, is larger by nearly a factor of two than the
amount of Mg-p character in MgCNi$_3$.

\begin{figure}[h]
\includegraphics[height=2.4in]{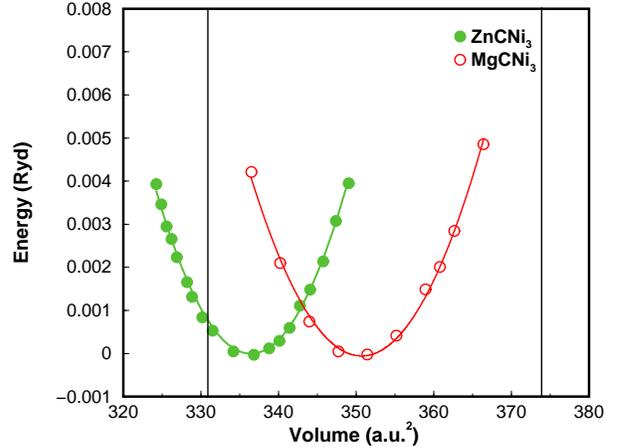}
\caption{The bulk moduli and equilibrium volumes of MgCNi$_3$ and
ZnCNi$_3$.  The experimental volumes are shown as vertical lines -
that of ZnCNi$_3$ is smaller than the calculated value.}
\end{figure}

The calculated equilibrium volume of ZnCNi$_3$ compares very favorably
with the reported value, the latter being 0.53\% smaller by lattice
constant. However, this result is somewhat unusual in that the
theoretical value is actually larger than the experimental one.  The
calculated equilibrium value of MgCNi$_3$ is 2\% smaller in lattice
constant than the experimental value.  The discrepancy is in the more
common direction, but larger than usual. The two energy vs. volume curves
are shown in Fig. 2, along with the experimental volumes. The bulk
modulus of ZnCNi$_3$ taken at the theoretical volume is 251 GPa whereas
that of MgCNi$_3$ taken at its own theoretical volume is 214 GPa.  The
MgCNi$_3$ result is similar to that obtained through LMTO calculations
\cite{TGK+02}.  Both these values are obtained by fitting the Murnaghan
equation of state \cite{FDM44} to an energy vs. volume curve and
extracting the bulk modulus through the relation: B = V$\frac{\partial^2
E}{\partial V^2}$.  The smaller volume Zn compound is, as expected,
harder than the Mg compound.  However, the bulk moduli of these two
compounds, calculated at their respective experimental volumes, differ by
only 3.6\%, with MgCNi$_3$ being {\it harder} than ZnCNi$_3$.  This is a
consequence of finding the theoretical equilibrium value of ZnCNi$_3$
above the reported value, while that of MgCNi$_3$ is below the reported
value.

\subsection{Electron-Phonon Coupling}
The empirical value quoted for the electron-phonon coupling
constant, $\lambda$, in MgCNi$_3$ depends on the method by which it is
obtained.  Using the size of the jump in specific heat at the
superconducting transition and assuming weak coupling BCS behavior
yields \cite{THQH+01} $\lambda$ = 0.79.  This method is obviously
unavailable for ZnCNi$_3$ since no transition has yet been observed.  
$\lambda$ can alternatively be derived by the more common method of
comparing experimental and theoretical results for the linear
coefficient $\gamma$ from specific heat data:

\begin{equation}
\lambda = \frac{\gamma_{exp}}{\gamma_{th}} - 1 ; \hspace{.1 in}  
\gamma_{th} 
= \frac{\pi^2k_{B}^2}{3}N(0) 
\end{equation}   

There is some variation in the reported values of $\gamma_{exp}$ for
MgCNi$_3$.  Some sources \cite{THQH+01, M-P+, ZQM+03} place the value 
at around
29 mJ/mol K$^2$ for the zero-field value, while others \cite{LShan03,
JYLin03} cite a higher value of about 33.5 mJ/mol K$^2$.  Using this
range of values, we obtain $\lambda$ = 1.5-1.75 for the Mg compound, in
agreement with previous results derived in this way.  \cite{DJS01,walte}
However, using this methodology, a negative $\lambda$ results for
ZnCNi$_3$, due to the small $\gamma_{exp}$=6.77 reported \cite{M-P+},
less than 25\% of that of MgCNi$_3$.  This unphysical result highlights the
discrepancy between experimental and theoretical comparisons of these
two compounds.  Large differences in observed specific heat data
combined with very small differences in calculated electronic structure
properties produce this unphysical value for $\lambda$.  

Furthermore, it is the ratio of the $\gamma$'s from the two different
compounds that stipulates that the DOS of the ZnCNi$_3$ sample must be
significantly lower than that of MgCNi$_3$.  Park {\it et al}
\cite{M-P+} use the definition in Eq. 1 along with their specific heat
data to put an upper bound on the value of the DOS of the Zn compound
relative to that of the Mg compound at the Fermi level.

\begin{equation} \frac{\gamma_{Mg}}{\gamma_{Zn}} = \frac{N(0)_{Mg}
(1+\lambda_{Mg})}{N(0)_{Zn} (1+\lambda_{Zn})} \end{equation}

It is clear from this equation that N(0)$_{Zn}$ takes its greatest
value when $\lambda_{Zn}$ = 0.  The ratio then yields N(0)$_{Zn}$
$\leq$ 0.41N(0)$_{Mg}$.  As mentioned above, there is no such large
depression of the ZnCNi$_3$ DOS as compared to the MgCNi$_3$ DOS at the 
Fermi level.  
In fact, our calculated value of N(0)$_{Zn}$ exceeds the derived upper
bound by almost a factor of two.

\begin{table}[t!]
\caption{Comparison of MgCNi$_3$ and ZnCNi$_3$
(Experimental values taken from Park \cite{M-P+} {\it et al})} 
\begin{center}
\begin{tabular}{|c||c|c|c|cr|} \hline

            &{ } N(0) ev$^{-1}$ &{ } $\gamma_{exp}$ mJ/mol$\cdot$K$^2$
&{ }$\Theta_D$ K &{ }
$\lambda$ & \\
\hline
MgCNi$_3$   &{ }5.003 & { }29.50 &{ } 255.9 &{ }1.5 &  \\
\hline
 ZnCNi$_3$  &{ } 4.049 & { }6.77  &{ } 421.3 &{ }-0.29&   \\ \hline

\end{tabular} 
\end{center}
\label{table1}
\end{table}

\section{Discussion} 

MgCNi$_3$ has now been studied fairly extensively, and we review some
results that may be relevant. According to Ren {\it et al.} \cite{Ren1},
the carbon occupancy of MgCNi$_3$ is sensitive to preparation
conditions and two different phases of the compound emerge.  The
$\alpha$-phase is carbon depleted, while the $\beta$ phase is nearly
stoichiometric (carbon occupancy is 0.96).  Both $\alpha$ and $\beta$
phases share the same cubic space group, but the $\alpha$ phase lattice
parameter is 1.3\% smaller and unlike the $\beta$-phase, it does not
superconduct.  This is consistent with previous studies \cite{TGA+02}
which found that T$_c$ decreases linearly with decreasing carbon
concentration until eventually, at a carbon occupancy of around 
0.88-0.89,
a multi-phase region is reached in which bulk superconductivity no
longer exists.  The reported $\alpha$-phase occured at an occupancy of
0.75 at the carbon site \cite{Ren1}, well within this multi-phase
region.  Shan {\it et al} \cite{LShan03} found that the 
specific heat $\gamma$ was 50\% lower in
the $\alpha$-phase than in the $\beta$-phase. The $\alpha$-phase can
then be distinguished from the $\beta$-phase in three important
aspects: it does not superconduct, it has a significantly smaller
$\gamma$, and it's equilibrium lattice constant is 1.3\% smaller.

Most if not all of the evidence regarding ZnCNi$_3$ can be reconciled if
we suppose that the phase reported by Park {\it et al} is a carbon
deficient ``$\alpha$ - ZnCNi$_3$'' phase corresponding to
$\alpha$-MgCNi$_3$.  The electronic structure of the two compounds is so
similar that it is reasonable to assume that carbon deficiencies in the
Zn structure would have much the same effect as carbon deficiencies in
the Mg structure.  Assuming that the experimental results for this
compound were taken from an $\alpha$-phase of ZnCNi$_3$, all
discrepancies between theory and experiment discussed in this paper
disappear.  A 1.3\% increase in the lattice parameter would result in the
common situation in which the theoretical value is smaller than the
experimental one.  If $\gamma$ is multiplied by a factor of two, as it
would be in moving from an $\alpha$ to $\beta$ phase, the $\lambda$ value
calculated using Eq. 1 has a value of 0.42, eliminating the non-physical
negative result.  Previous electronic structure calculations show that in
MgCNi$_3$, N(0) decreases dramatically as C concentration
decreases\cite{LShan03}, resulting in suppression of the superconducting
transition. Similar effects would be expected for ZnCNi$_3$.

Another striking difference in the data
between the Zn and Mg based compounds is a
sharply increased lattice stiffness $\Theta_D$.  
Even in the minimally doped alloy
Mg$_{0.85}$Zn$_{0.15}$CNi$_3$, an increase of 38\% was observed
\cite{park1} for $\Theta_D$ and in the fully Zn substituted compound, the
increase is 67\% \cite{M-P+}. The addition of Zn in any concentration
causes a volume contraction and a concurrent hardening of phonon modes in
general. In pure MgCNi$_3$, the frequency of a very soft acoustic
Ni-based phonon mode is calculated in the harmonic approximation to
become negative along much of the $\Gamma-M$ direction of the BZ
\cite{sav,heid}. Anharmonic stabilization of this mode results in
observed dynamic displacements of the Ni ions perpendicular to the Ni-C
direction. \cite{heid}. This ``breathing"  distortion allows each Ni ion
to move away from its two C neighbors and toward the empty interstitial
site.  In a C deficient compound, stress on the Ni ions would be
partially relieved by vacancies, reducing the advantage of such
distortions and thereby increasing $\Theta_D$.

\section{Conclusions} From our calculations, stoichiometric ZnCNi$_3$ and
MgCN$_3$ are very much alike in both structural and electronic
properties.  The experimental report of widely differing specific heat
data and the lack of superconductivity down to 2 K seems highly unusual
in light of the close similarity of these two compounds.  The rather
large suppression of the DOS at the Fermi energy required to interpret
the experimental results using BCS theory fails to materialize from the
calculations.  All results are in line with a ZnCNi$_3$ phase that is
carbon-deficient rather than stoichiometric.  Carbon deficient MgCNi$_3$
is known to have a smaller volume than the stoichiometric compound, to
have a strongly depressed $\gamma$, and to be non-superconducting.  
Our results suggest that the lattice constant of stoichiometric ZnCNi$_3$
is likely to be larger than that which is reported (probably near 3.74
$\AA$), and that a depression of less than 20\% in N(0) occurs.  A truly
stoichiometric ZnCNi$_3$ compound would likely be superconducting at only
a somewhat lower temperature than MgCNi$_3$.

\bibliography{joh}

\end{document}